\documentclass{PoS}

\usepackage{ amssymb }

\title{Study of the scalar charmed-strange meson $D_{s0}^*(2317)$ with chiral fermions}

\ShortTitle{$D_{s0}^*(2317)$ with chiral fermions}

\author{\speaker{M. Gong}\thanks{For the $\chi$QCD Collaboration.}\\
        Dept. of Physics and Astronomy, University of Kentucky, Lexington, KY 40506, USA\\
        E-mail: \email{gongming@pa.uky.edu}}

\author{A. Li\\
        Dept. of Physics, Duke University, Durham, NC 27708, USA\\
        E-mail: \email{anyili@phy.duke.edu}}

\author{A. Alexandru\\
        Dept. of Physics, George Washington University, Washington, DC 20052, USA\\
        E-mail: \email{aalexan@gwu.edu}}

\author{Y. Chen\\
        Institute of High Energy Physics, Chinese Academy of Science, Beijing 100049, China\\
        E-mail: \email{cheny@ihep.ac.cn}}

\author{T. Draper\\
        Dept. of Physics and Astronomy, University of Kentucky, Lexington, KY 40506, USA\\
        E-mail: \email{draper@pa.uky.edu}}

\author{K.F. Liu\\
        Dept. of Physics and Astronomy, University of Kentucky, Lexington, KY 40506, USA\\
        E-mail: \email{liu@pa.uky.edu}}

\abstract{
The recently discovered charmed-strange meson $D_{s0}^*(2317)$ has
been speculated to be a tetraquark mesonium. We study this suggestion with overlap fermions
on 2+1 flavor domain wall fermion configurations. We use 4-quark interpolating operators 
with $Z_4$ grid sources on two lattices ($16^3 \times 32$ and $24^3 \times 64$) to study
the volume dependence of the states in an attempt to discern the nature of the states in
the four-quark correlator to see if they are all two-meson scattering states or if one is a tetraquark mesonium.
We also use the hybrid boundary condition method for this purpose which is designed to lift 
the two-meson states in energy while leaving the tetraquark mesonium unchanged. We find
that the volume method is not effective in the present case due to the fact that
the scattering states spectrum is closely packed for such heavy states so that
one cannot separate out individual scattering states since the volume dependence is skewed
as a result. However, the hybrid boundary condition method works and we found that
the four-quark correlators can be fitted with a tower of two-meson scattering states.
We conclude that we do not see a tetraquark mesonium in the 
$D_{s0}^*(2317)$ meson region.
}

\FullConference{The XXVIII International Symposium on Lattice Field Theory, Lattice2010\\
		June 14-19, 2010\\
		Villasimius, Italy}

\begin{document}

\section{Introduction}

A charmed-strange meson $D_{s0}^*(2317)$ has been found in recent years~\cite{babar,cleo}.
Its mass is $2317.8\pm0.6$ MeV, which is significantly lower than the one predicted from the quark model~\cite{quarkmodel1,quarkmodel2}.
This discrepancy has prompted speculations that $D_{s0}^*(2317)$ is a DK molecule~\cite{DK}, a four quark state~\cite{4quark}, 
or a dynamically generated bound $c\bar{s}$ state through the coupling to the nearby DK threshold~\cite{threshold}.

Lattice QCD is a powerful tool used to calculate the hadron spectrum from first principles.
Some earlier works were done with dynamical lattice QCD for the $c\bar{s}$ meson~\cite{lattice1,lattice2,lattice3,lattice4}, but the puzzle is not 
settled since the results are significantly different than the experimental mass.
A recent calculation~\cite{dong09} with overlap fermion on $2+1$ flavor domain-wall sea gives a mass of $2304\pm22$ MeV which is 
consistent with the experimental value.
Presumably, this is largely due to the fact that the overlap fermion has much smaller $O(m^2a^2)$ error than the other fermion formulations and is suitable for both light and charm quarks~\cite{dong07}.
This result suggests that the $D_{s0}^*(2317)$ is just the scalar $c\bar{s}$ meson.

In this article, we shall study this meson to see if it can be a tetraquark mesonium
as suggested in the literature~\cite{DK,4quark}.
We calculate the spectrum with 4-quark interpolating operators and look for evidence of a 4-quark state.

\section{Strategies for the computation}

\subsection{Overlap valence fermion on domain-wall sea}

The overlap fermion action obeys chiral symmetry at finite lattice spacing and is, thus, free of $O(a)$ errors. 
It is shown that the effective quark propagator of the massive overlap fermion has the same form as that of the continuum~\cite{continuum}.
The $O(m^2a^2)$ error, which is important in the charm region, is estimated to be small on quench lattices~\cite{quench1,dong07} and even 
smaller on the dynamical domain-wall sea~\cite{dong09} due to HYP smearing.
Thus, it is shown that it can be used for both charm and light quarks on the $24^3 \times 64$ lattice DWF configurations~\cite{li10}.

We use the valence overlap fermion on $N_f = 2+1$ domain wall dynamical configurations from the RBC and UKQCD collaborations~\cite{rbc}.
This is a mixed action approach.
Since the valence is a chiral fermion, only one extra low-energy constant $\Delta_{mix}$ needs to be determined~\cite{deltamix1,deltamix2}, which turns out to be small~\cite{li10}.

\subsection{The $Z_4$ grid source}

We introduce $Z_4$ grid sources to gain more efficiency.
A $Z_4$ grid source is defined as:
\begin{equation}
\eta(\vec{x}) = \sum_{\vec{i}\in \mathcal{G}} \theta(\vec{i}) \delta_{\vec{x},\vec{i}}
\end{equation}
where $\mathcal{G}$ is a sparse grid of lattice sites on timeslice $t=0$, and $\theta(\vec{i}) \in \{1,i,-1,-i\}$ is the $Z_4$ random phase on site $\vec{i}$.

The corresponding quark propagator is:
\begin{eqnarray}
G(\vec{y},\eta) &=& D^{-1}(\vec{y},\vec{x}) \eta(\vec{x}) \nonumber\\
       &=& \sum_{\vec{i}\in \mathcal{G}} \theta(\vec{i}) D^{-1}(\vec{y},\vec{x}) \delta_{\vec{x},\vec{i}} = \sum_{\vec{i}\in \mathcal{G}} \theta(\vec{i}) G(\vec{y},\vec{i})
\end{eqnarray}

And the anti-quark propagator is:
\begin{eqnarray}
G(\eta,\vec{y}) &=& \left( D^{\dagger -1}(\vec{y},\vec{x}) \eta^*(\vec{x}) \right)^{*,T_d, T_c} \nonumber\\
		&=& \sum_{\vec{i}\in \mathcal{G}} \theta(\vec{i}) G(\vec{i},\vec{y})
\end{eqnarray}
where the $T_d$ and $T_c$ are transpose operations on dirac space and color space respectively.

On space-time dimensions with implicit color and dirac indices, all connected 4-quark correlation functions have the form:
\begin{eqnarray}
C(\vec{y},\eta) &=& Tr \left< \Gamma_1 G_1(\vec{y},\eta) \Gamma_2 G_2(\vec{y},\eta) \Gamma_3 G_3(\eta,\vec{y}) \Gamma_4 G_4(\eta,\vec{y}) \right> \nonumber\\
		&=& \sum_{\vec{i_1},\vec{i_2},\vec{i_3},\vec{i_4}\in \mathcal{G}} \theta(\vec{i_1}) \theta(\vec{i_2}) \theta(\vec{i_3}) \theta(\vec{i_4}) Tr \left< \Gamma_1 G_1(\vec{y},\vec{i_1}) \Gamma_2 G_2(\vec{y},\vec{i_2}) \Gamma_3 G_3(\vec{i_3},\vec{y}) \Gamma_4 G_4(\vec{i_4},\vec{y}) \right> \nonumber\\
		&=& \sum_{\vec{i}\in \mathcal{G}} Tr \left< \Gamma_1 G_1(\vec{y},\vec{i}) \Gamma_2 G_2(\vec{y},\vec{i}) \Gamma_3 G_3(\vec{i},\vec{y}) \Gamma_4 G_4(\vec{i},\vec{y}) \right> \nonumber\\
		&& + \sum_{\vec{i_1},\vec{i_2},\vec{i_3},\vec{i_4} {{\in \mathcal{G}} \atop{not \,equal}}} \theta(\vec{i_1}) \theta(\vec{i_2}) \theta(\vec{i_3}) \theta(\vec{i_4}) Tr \left< \Gamma_1 G_1(\vec{y},\vec{i_1}) \Gamma_2 G_2(\vec{y},\vec{i_2}) \Gamma_3 G_3(\vec{i_3},\vec{y}) \Gamma_4 G_4(\vec{i_4},\vec{y}) \right> \nonumber\\
&&
\label{eq:grid}
\end{eqnarray}
where the $\Gamma$'s are $\gamma$ matrices.

\begin{figure}
\begin{center}
\includegraphics[width=2in]{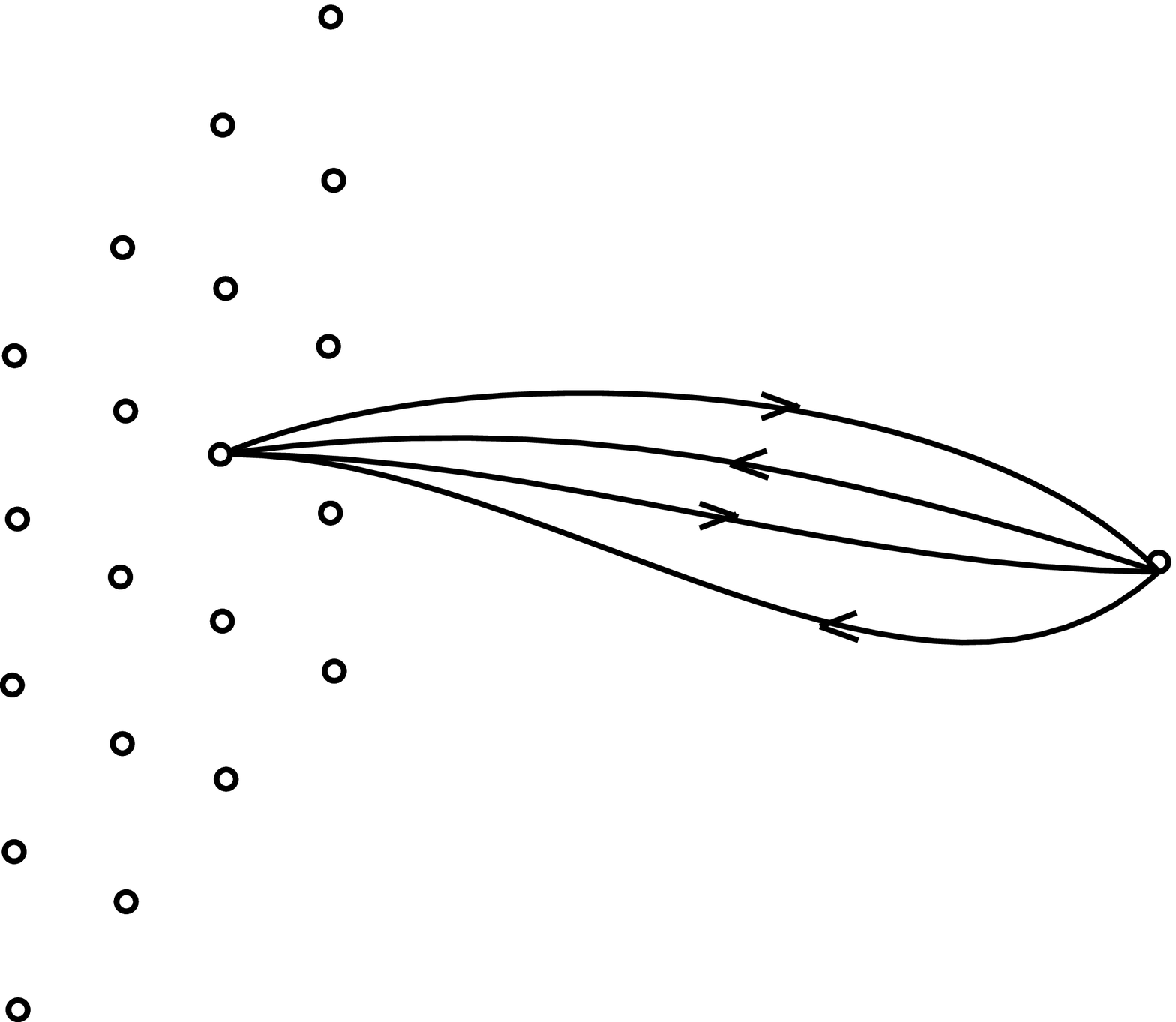}
\hspace{2cm}
\includegraphics[width=2in]{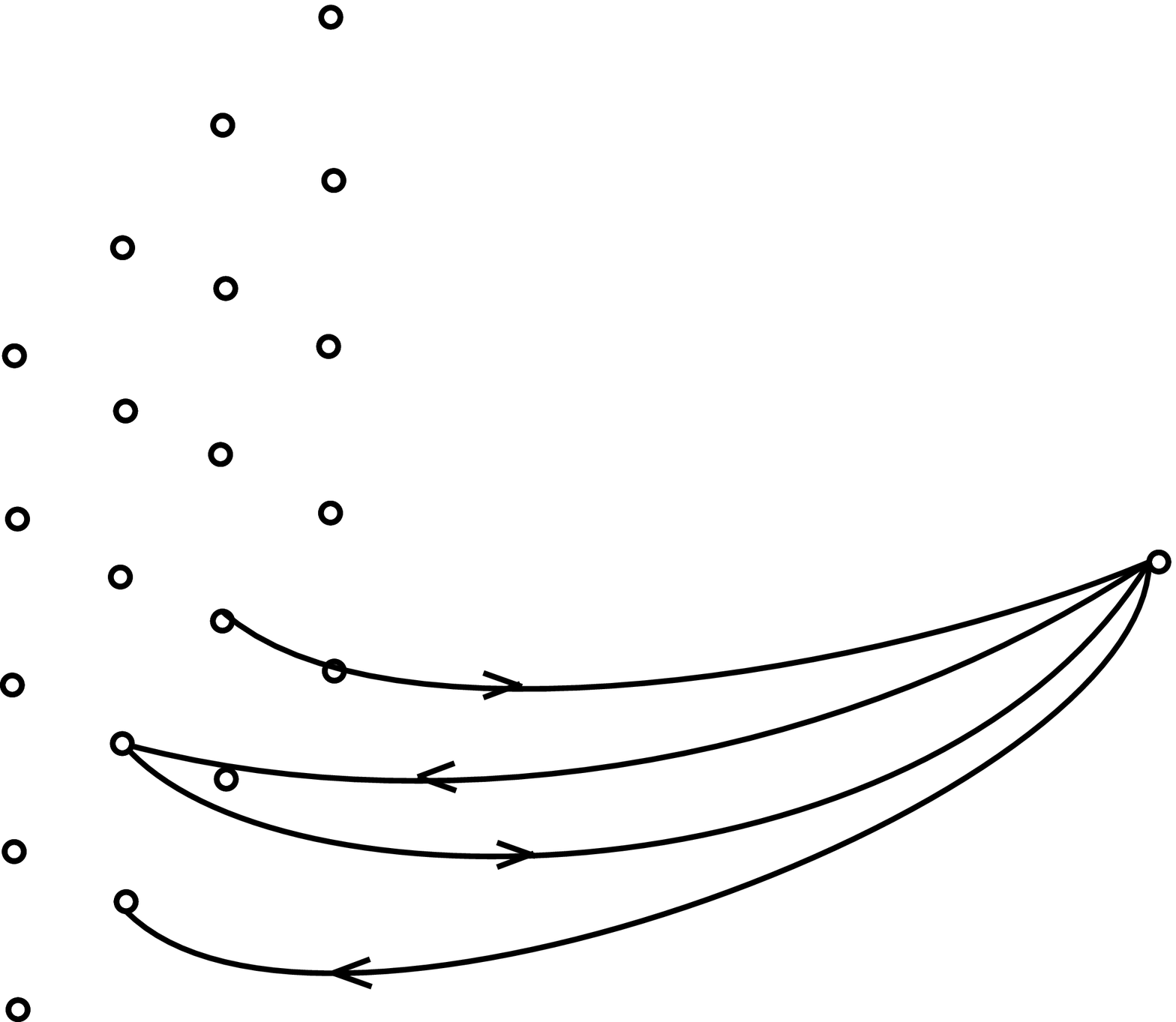}
\end{center}
\caption{
\label{fig:z4grid} A sketch of the correlators with $Z_4$ grid 
sources. a) Left panel represents the first term 
in Eq.~(\protect \ref{eq:grid}) which is a signal. b) Right 
panel represents the second term in Eq.~(\protect \ref{eq:grid}) which is a noise.
}
\end{figure}

The second term is stochastically eliminated with many gauge configurations and/or many noise sources.
In this limit, the correlator is a good approximation of the sum of many correlators with point sources.
Figure \ref{fig:z4grid} shows the different diagrams of the two terms.

\subsection{Hybrid spatial boundary condition}

Both 2-meson states and the possible 4-quark state are present in Eq.~(\ref{eq:grid}) and it is difficult to distinguish them, so we adopt the hybrid spatial boundary condition to help in this regard~\cite{hybridbc}.

If we use the periodic spatial boundary condition, the available momenta on quark propagators are in the set
$P = \{ 0, \pm \frac{2\pi}{L}, \pm \frac{4\pi}{L}, \dots \}$.
On the other hand, if we use the anti-periodic spatial boundary condition, the available momenta on quark propagators are in the set
$A = \{ \pm \frac{\pi}{L}, \pm \frac{3\pi}{L}, \dots \}$.
We calculate the quark propagators with the periodic spatial boundary condition while calculate the anti-quark propagators with the anti-periodic spatial boundary condition.
With this hybrid spatial boundary condition, the momenta of mesons are each in the set $P + A = A$ while the momenta of 4-quark state are in the set $2 P + 2 A = P$.
The lowest momentum of a 4-quark state is $(0,0,0)$, while the lowest momentum of a $q \bar q$ meson state is lifted to $(\pm \frac{\pi}{L},\pm \frac{\pi}{L},\pm \frac{\pi}{L})$.

The $D_{s0}^*(2317)$ state is about $45$ MeV lower than the $D K$ threshold. 
If $D_{s0}^*(2317)$ is a tetraquark mesonium, it would remain at its position when the hybrid boundary condition is
imposed as compared to the periodic condition. 
On the other hand, the $D K$ threshold will be shifted up by $177$ MeV on the $24^3\times 64$ lattice, making it easier to identify the tetraquark mesonium state.

\section{Simulation results}

\subsection{Simulation parameters}

We use the $N_f=2+1$ domain wall dynamical configurations from RBC and UKQCD collaborations.
The lattice sizes are $16^3\times 32$ and $24^3\times 64$ with lattice spacing $1/a = 1.73(3)$ GeV.
For each ensemble, we use $37$ configurations.

The propagators are calculated with the multimass algorithm with masses ranging from 0.00140 to 0.70.
For this case, the mass of $u/d$ quarks is set to 0.0135, the corresponding $m_\pi$ is $310$ MeV which is near the $u/d$ quark mass in the sea. The strange quark mass is set to $0.067$ and the charm quark mass is set to $0.68$.
The grid spacing of the source is set to $8$ on spatial dimensions. We have one $Z_4$ grid source per configuration.

\subsection{The interpolating operators and possible states}

The 4-quark interpolating operator we use is $\bar{c}^\alpha \gamma_5 u^\alpha \bar{u}^\beta \gamma_5 s^\beta$.
It has the quantum number of $D_{s0}^*(2317)$ except with isospin $I=1$.

The correlation functions contain $D K$ scattering states, $D_s \pi$ scattering states, a possible tetraquark state, and other higher states.

The data are not good enough to determine the ratio of the spectral weights of the $D K$ scattering states and the $D_s \pi$ scattering states exactly and the fitting results suggest that the spectral weights of $D K$ states and the $D_s \pi$ states are of the same order.
Therefore, we assume that all the scattering states have the same spectral weight.
We propose to check the assumption carefully with open-jaw diagram calculations in the future.

The length of the time dimension of the $16^3\times 32$ lattice is relatively small, and therefore the wrap-around states cannot be neglected.
If the meson-meson interaction is neglected, the correlation functions of the wrap-around states have the form:
\begin{equation}
C(t) = W \left( e^{-E_1 t -E_2 (T-t)} + e^{-E_2 t -E_1 (T-t)} \right)
\end{equation}
In this case, the spectral weights of the forward propagating 2-meson scattering states and the wrap-around states are the same.
Considering the states and the assumptions mentioned above, we construct a model for the fitting of our data:
\begin{equation}
C(t) =  C_{D_s \pi}(t) + C_{D K}(t) + C_{D_s \pi}^{wrap}(t) + C_{D K}^{wrap}(t) + C_{T}(t)
\end{equation}
In this model, the 2-meson scattering states, the 2-meson wrap-around states, and the 4-quark state are:
\begin{eqnarray}
C_{A B} &=& W \sum_{p\in P/A} \frac{e^{- \left(E_A(p)+E_B(-p) + \Delta E\right) t}}{E_A(p) E_B(-p)} + (t \leftrightarrow T-t)\label{eq:model1}\\
C_{A B}^{wrap} &=& W \sum_{p\in P/A} \frac{e^{- E_A(p) t -E_B(-p) (T-t)}}{E_A(p) E_B(-p)} + (t \leftrightarrow T-t) \label{eq:model2}\\
C_{T}(t) &=& W^\prime e^{- E^\prime t} \label{eq:model3}
\end{eqnarray}
where the "$A B$" can be "$D K$" or "$D_s \pi$" and the particle energy of $A$ or $B$ is $E_{A/B}(p) = \sqrt{m_{A/B}^2 + p^2}$.
The momentum $p$ is summed over the corresponding set of momenta.
We have calculated with both the original periodic spatial boundary condition and with the hybrid spatial boundary condition.

In the model, the $q\bar{q}$ masses are input from fits of the single meson correlation functions.
We have four parameters to fit in the model.
The $W$ in Eqs.~(\ref{eq:model1}) and (\ref{eq:model2}) is the spectral weight of all the scattering states and wrap-around states, the $\Delta E$ in Eq.~(\ref{eq:model1}) is the mean effective energy of meson-meson interactions, and $E^\prime$ and $W^\prime$ in Eq.~(\ref{eq:model3}) are the mass and the spectral weight of a possible tetraquark state respectively.

\subsection{Fitted results}

\begin{table}
\caption{\label{tab:result}Fitted results}
\begin{tabular}{|c|c|c|c|c|c|}
\hline
Lattice & $\chi^2/d.o.f.$ & $\Delta E a$ & $W$ & $E^\prime a$ & $W^\prime$ \\
\hline
$16^3\times32$ Periodic	& $0.463$	& $0.0040(64)$	& $1.07(6)\times10^{-7}$	& $2.26(3)$	& $1.83(24)\times10^{-4}$	\\
$16^3\times32$ Hybrid	& $1.074$	& $0.0037(64)$	& $2.02(1)\times10^{-7}$	& $2.23(9)$	& $1.71(82)\times10^{-4}$	\\
$24^3\times64$ Periodic	& $1.264$	& $-0.0019(40)$	& $1.74(1)\times10^{-7}$	& $2.18(3)$	& $2.83(32)\times10^{-4}$	\\
$24^3\times64$ Hybrid	& $0.946$	& $-0.0072(41)$	& $1.87(9)\times10^{-7}$	& $2.27(9)$	& $3.62(166)\times10^{-4}$	\\
\hline
\end{tabular}
\end{table}

The fitted results are tabulated in Table \ref{tab:result}.

We note first that $\Delta E$ is consistent with zero, which suggests the meson-meson interactions do not play an important role and can be neglected.
We note that on the $24^3\times 64$ lattices, the $W$ fitted from hybrid boundary condition coincides with that from periodic boundary condition. This confirms that our assumption on the fitting model are reasonable.
On the $16^3 \times 32$ lattices, the $W$ fitted from periodic and hybrid boundary condition do not agree with each other. We suspects that it is due to the finite volume effect with smaller time extent.

The fitted $E^\prime$ are all larger than $3.7$ GeV. We tried to fit it with a shallow constraint around $2.3 \backsim 2.4$ GeV, but we cannot find a state there. 
Therefore, the fitted $E^\prime$ shows the effects of higher states and there is no 4-quark state near $D_{s0}^*(2317)$.

\subsection{The volume dependence}

Another method~\cite{vlmdpd} to distinguish the tetraquark mesonium and the scattering states is the volume dependence.
The spectral weight of a single-particle state has little volume dependence while that for a two-particle scattering state exhibits a $1/L^3$ behavior. 

\begin{figure}
\begin{center}
\includegraphics[width=2.5in]{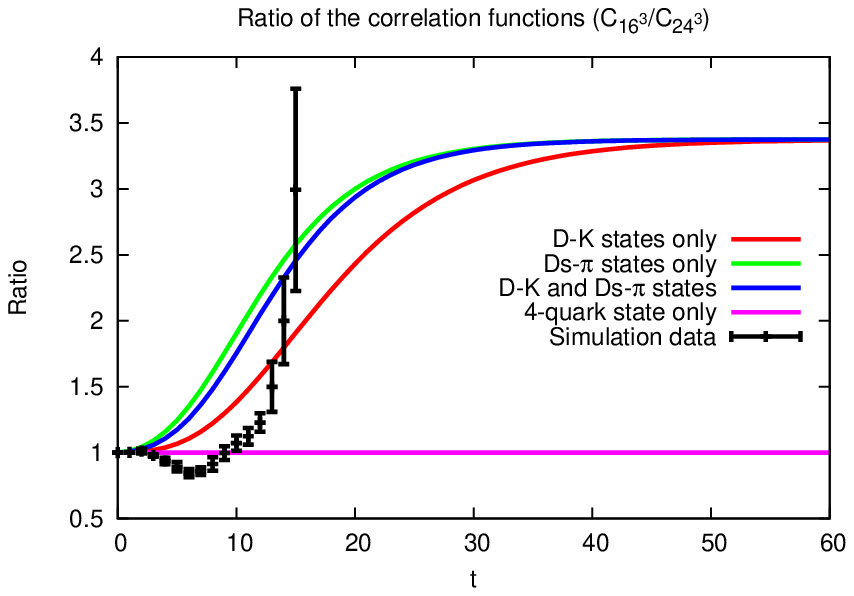}
\includegraphics[width=2.5in]{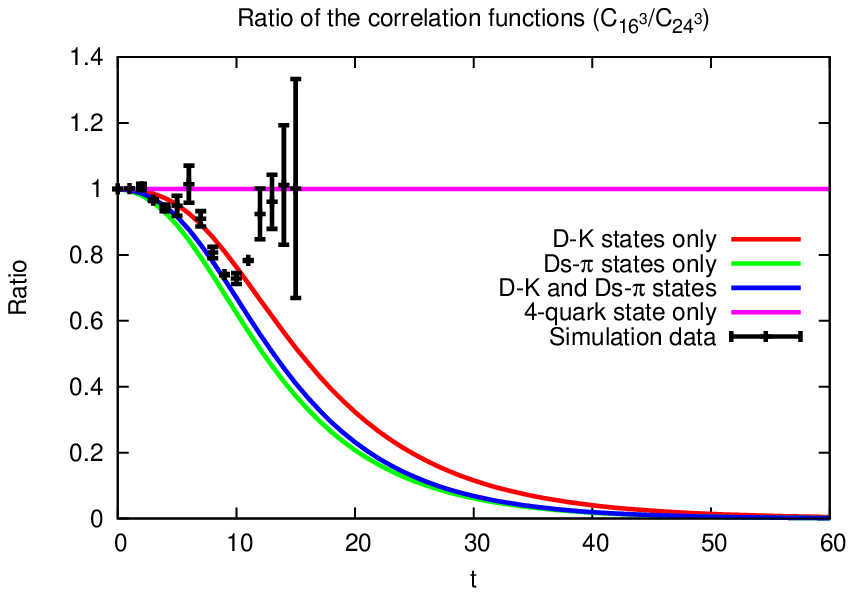}
\end{center}
\caption{\label{fig:ratio} The ratio of the correlation functions ($C_{16^3}/C_{24^3}$) with periodic/hybrid boundary condition.}
\end{figure}

We have tried this method and find that it is not a good criterion for our case.
In the charm region, there are towers of scattering states with small energy gaps.
We cannot separate the states and the volume dependence of the correlation functions becomes very complex.
Figure \ref{fig:ratio} shows the volume dependent behaviors of many states.
The horizontal axis is time and the vertical axis is the ratio of the correlation functions on different volumes.
The lines show the behaviors of a tower of $D K$ states, a tower of $D_s \pi$ states, the sum of $D K$ states and $D_s \pi$ states, and a tetraquark state only.
The points with error-bars on the plots show our real data.

For periodic boundary condition case, the tetraquark state should have the ratio 1 on all time slices while a scattering state should have a ratio at 3.4.
If there are many states in the correlation functions, the ratios run from 1 to 3.4 while the time increases.
It shows that we need a long time extent to observe the $1/L^3$ behavior of scattering states.
For the hybrid boundary condition case, the conclusion is similar.

The real data, however, contains $D K$ states, $D_s \pi$ states, other scattering states and excited states.
Therefore, the behavior is more complex and it is difficult to find a simple model to fit it.
We conclude that the time extent of our lattice is not long enough to resolve individual scattering state. This renders useless
the volume study to discern the number of particles of the state.

It is claimed in the lattice calculation~\cite{chiu} that tetraquark mesoniums are found on $(1.8)^3 \times 3.5$ fm and $(2.1)^3 \times 4.2$ fm lattices through the volume study.
The time extent is about the same as that of our lattice. 
In view of this study, the time extent in Ref.~\cite{chiu} is similarly shorter than
needed to resolve the tower the scattering states and, thus, suffers the same shortcoming as revealed in the present study.

\section{Conclusions}

We have studied the scalar charmed-strange meson with chiral fermions.
The correlation functions with the 4-quark interpolating operator are fitted and no 4-quark state is found in our data.
The result shows that the meson $D_{s0}^*(2317)$ is not likely a 4-quark state.

The fact that $D_{s0}^*(2317)$ mass is well reproduced as a $c\bar{s}$ meson in the study with chiral fermions~\cite{dong09}, the present calculation reinforces this interpretation.

We find that the volume method is not effective in the present case due to the fact that the scattering states spectrum are closely packed for such heavy states and one cannot separate out individual scattering states and its volume dependence is skewed as a result. 

The open-jaw diagram will be studied in the future. 
More information about the spectral weights can be extracted in the study and we can thus make a more precise fitting model.

The $I=0$ channel will be studied with disconnected diagrams and variational method.

With similar methods, other interesting states in the charm region can be studied, such as $X(3872)$, $Z^+(4430)$, etc..

\end{document}